\documentclass[apl,aip,twocolumn,reprint,preprintnumbers,amsmath,amssymb]{revtex4-1}

\usepackage{color}
\usepackage{graphicx}
\usepackage{dcolumn}
\usepackage{bm}

\begin{document}

\title{A report on the magnetic order of transition-metal $\delta$-doped cubic ZnO}

\author{I. Galanakis}\email{galanakis@upatras.gr}
\affiliation{Department of Materials Science, School of Natural
Sciences, University of Patras,  GR-26504 Patra, Greece}

\begin{abstract}
Preliminary results on the properties of transition-metal (Ti, V,
Cr, Mn, Fe, Co, Ni) $\delta$-doped ZnO are reported. Using
\emph{ab-initio} electronic structure calculations the magnetic
order is studied assuming both the cubic rock-salt and zinc-blende
structures for ZnO. The ground state magnetic order is found to
depend strongly on the transition-metal atom.
\end{abstract}

\maketitle

ZnO is a well-known wide-band semiconductor crystallizing in the
wurtzite (WZ) structure.\cite{Rev1} When grown as a thin film the
lattice structure adopted is the cubic analog the zinc-blende (ZB)
structure, while a pressure of about 6 GPa induces the cubic
rock-salt (RS) structure.\cite{Rev1} Both cubic structures have
similar lattice constants (4.47 \AA\ for the ZB and 4.25\AA\ for
the RS). Both can be viewed in general as a fcc lattice with four
sites as basis distributed along the diagonal in equidistant
positions. In both structure the Zn atoms occupy the origin of the
unit. In the RS the oxygen atoms are located at the middle of the
diagonal and in reality they bridge the Zn atoms being positioned
in the middle of the line connecting the nearest-neighboring Zn
atoms being located in the same (001) layer with them . In the ZB
structure the oxygen atoms occupy the second site along the
diagonal and they create a pure O (001) layer which lies
in-between two successive pure (001) Zn layers. Finally we should
remark that the WZ and RS structures are similar since the Zn and
O atoms show similar electronic properties in both structures;
only the width of the bands in the density of states (DOS) is
somewhat smaller in the WZ structure.

ZnO has started to attract considerable consideration in
spintronic/magnetoelectronic research since it was discovered that
the occurrence of Zn/O antisites or defects can lead to the
appearance of magnetism.\cite{Rev2} Moreover the case of
transition-metal impurities has been widely studied\cite{Sato}
(especially Co and Mn impurities) but contradictory results are
found in literature both within the experimental and theoretical
articles and thus the appearance of magnetic order in ZnO diluted
magnetic semiconductors still remains an open
issue.\cite{Andriotis,Iusan,Petit,Zunger} In this repot we present
preliminary results on the transition-metal $\delta$-doped cubic
ZnO structures. We have considered a tetragonal
1$\times$1$\times$2 unit cell and along the (001) direction we
have substituted one Zn layer with a transition-metal layer; thus
if we neglect the oxygens we have one transition-metal layer
followed by three Zn layers which is repeated along the (001)
direction. Within the transition metal layer we have considered
both ferromagnetic (FM) and antiferromagnetic (AFM) coupling of
the nearest-neighboring transition-metal atoms and we have
calculated the total energies of the considered magnetic
configuration. To perform the calculations we have employed the
full-potential nonorthogonal local-orbital minimum-basis band
structure scheme (FPLO)\cite{koepernik} within the generalized
gradient approximation (GGA) in the Perdew-Burke-Ernzerhof
formulation.\cite{gga} Moreover in the reciprocal \textbf{k}-space
we considered a 12$\times$12$\times$6 grid in the Brillouin zone
to perform the integrals calculations. The Zn 3\textit{d} states
were treated as valence states and they result in a narrow band
low in energy with respect to the other valence states.

\begin{table}
\caption{Calculated atom-resolved spin magnetic moments (in
$\mu_B$) for the transition-metal atoms for the two cubic
structure under study (ZB and RS) and for the cases of magnetic
coupling taken into account. In cases where dashed lines are
present, we were not able to converge the self-consistent
calculations.}
\begin{ruledtabular}
\begin{tabular}{l|cc|cc}
$m^X$ & \multicolumn{2}{c|}{ZB} & \multicolumn{2}{c}{RS} \\
\hline

$X=$ & AFM & FM & AFM & FM\\ \hline

Ti&  0.000   &-----&   1.163  & -----\\

V &  2.027   &2.736 &  2.668 &2.615\\

Cr & 3.237   &-----  & 3.787  & 3.743\\

Mn  &-----   &4.701   &----- &4.640\\

Fe  &3.523  & 3.759   &3.751 &  -----\\

Co&  2.474 &  2.655   &2.515& -----\\

Ni  &1.308&   1.575   &1.658& ----- \\
\end{tabular}
\end{ruledtabular}
\label{table1}
\end{table}

\begin{table}
\caption{Energy difference in eV. 2nd and 3rd column are the
differences between the ZB and the RS ($\Delta
\mathrm{E}_{ZB-RS}$) structure in both the FM and AFM
configurations; positive value means that the RS structure is
favored while negative means that the ZB is favored. 4th and 5th
column are the differences between the AFM and the FM ($\Delta
\mathrm{E}_{AFM-FM}$) orderings in both the ZB and RS structures;
positive value means that the FM configuration is favored while
negative means that the AFM is favored. The last column denotes
which is the stable ground state. Dashes correspond to the case
where the $\Delta \mathrm{E}$s cannot be computed since they
involve not-converged calculations.}
\begin{ruledtabular}
\begin{tabular}{l|cc|cc|c}

& \multicolumn{2}{c|}{$\Delta \mathrm{E}_{ZB-RS}$} & \multicolumn{2}{c|}{$\Delta \mathrm{E}_{AFM-FM}$} & Stable\\

$X=$ & AFM & FM & ZB & RS & Ground State\\ \hline

Ti  &+0.511&  -----& ----- &  -----   &AFM-RS\\

V   &+0.933 & +0.109&  +0.425&  -0.398 & AFM-RS\\

Cr & +0.474  &----- & -----  & -0.304  &AFM-RS\\

Mn& -----  & -0.624  &-----  & -----   &FM-ZB\\

Fe  &-1.342&  -----  & -0.242&  ----- &  AFM-ZB\\

Co  &-1.601&  ----- &-0.204 &-----   &AFM-ZB\\

Ni &  -0.068&  -----&   +0.380&  ----- &  FM-ZB\\

\end{tabular}
\end{ruledtabular}
\label{table2}
\end{table}

\begin{figure}[t]
\includegraphics[width=\columnwidth]{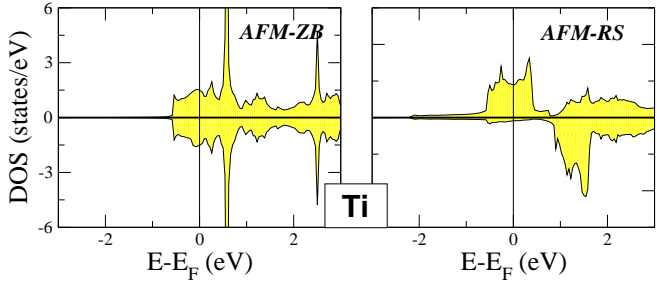}
 \caption{Density of state (DOS) of the Ti atom in
the various cases under study. Note that the Fermi energy is set
to zero. Positive (negative) DOS values correspond to the spin-up
(spin-down) states. In the AFM cases the DOS of the second Ti atom
has the two spin-bands exchanged.} \label{fig1}
\end{figure}

In table \ref{table1} we have gathered the spin magnetic moments
of the transition metal atoms for both ZB and RS structures. We do
not present the induced spin magnetic moments for the Zn and O
atoms since they are almost negligible with respect to the
transition metal atoms. Wherever no number is given, are the cases
where convergency could not be achieved. For Ti we were able to
converge only the AFM configurations but only in the RS structure
magnetism was present. As shown in Fig. \ref{fig1}, the AFM-ZB  is
a usual metal while in the AFM-RS case around the Fermi level we
get only spin-up states and the compound is in reality a
half-metal since the Zn and O atoms (not presented here) show an
almost semiconducting behavior. The later is also the ground state
as can be deduced from Table \ref{table2} where we present the
energy differences between the various configurations.

\begin{figure}[t]
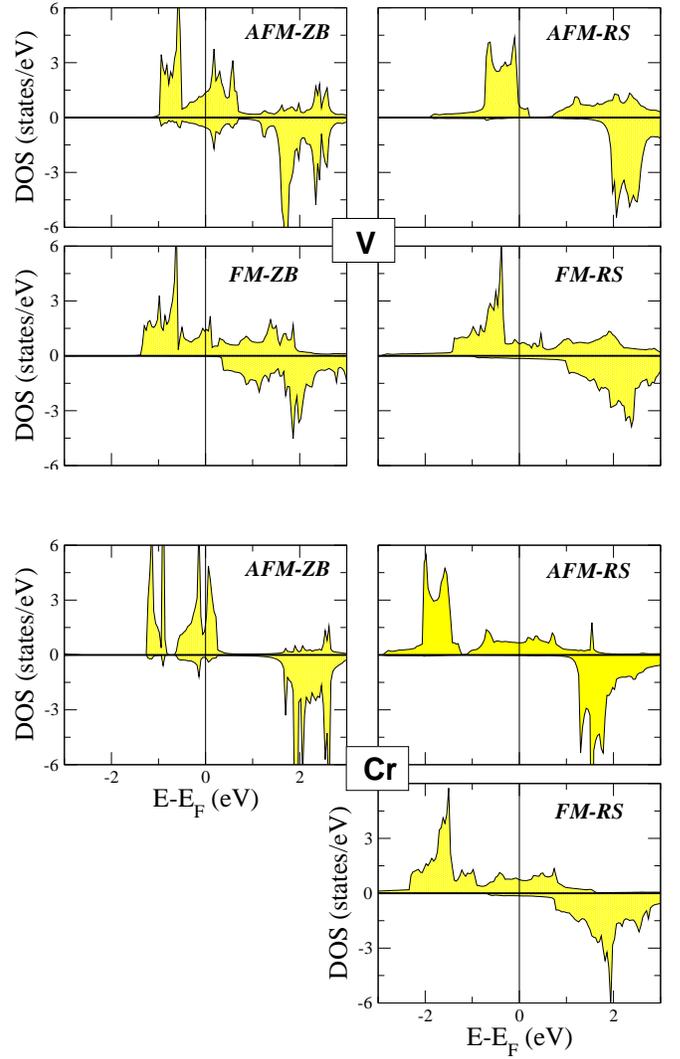

\includegraphics[width=\columnwidth]{fig2a.eps}
\vskip 0.2cm
\includegraphics[width=\columnwidth]{fig2b.eps}
\vskip -0.2 cm \caption{Similar to Fig. \ref{fig1} for the V and
Cr $\delta$-doped systems under study.} \label{fig2}
\end{figure}

When the $\delta$-doping is of vanadium character we were able to
converge all cases as shown in Table \ref{table1} with large spin
magnetic moments which approach the ideal 3$\mu_B$ of the isolated
single V impurity with the exception of AFM-ZB. The later case as
shown by the DOS is Fig. \ref{fig2} is the only case where a small
weight around the Fermi level is also present in the spin-down
band structure.  The most stable structure as can be deduced from
Table \ref{table2} is the AFM-RS structure as for the Ti-case,
which should be robust since the Fermi level is located after a
high-intensity pick in the spin-up DOS crossing a low of the
spin-up DOS. Cr shows similar behavior to V although we were not
able to converge the FM-ZB case. Again for the two magnetic
orderings of the RS we get spin magnetic moments approaching the
ideal 4$\mu_B$ of the isolated single Cr impurity and the AFM-RS
is the ground state as shown in Table \ref{table2}. DOS in Fig.
\ref{fig2} shows similar trends to the V one although almost
states corresponding to the one extra electron are occupied in the
spin-up DOS.

\begin{figure}[t]
\vskip 1cm
\includegraphics[width=\columnwidth]{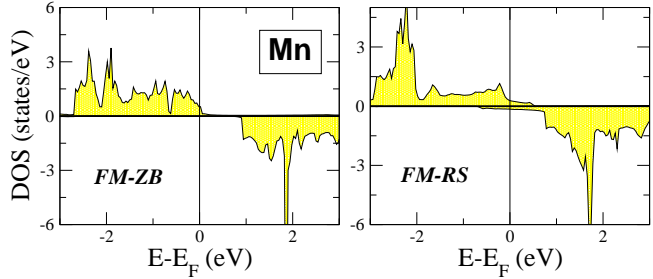}
\vskip -0.2 cm \caption{Similar to Fig. \ref{fig1} for the Mn
$\delta$-doped system under study.} \label{fig3}
\end{figure}

In the case of Mn, we were able to converge only the FM solutions
for both RS and ZB structures with very large Mn atomic spin
magnetic moments; they approach the 5 $\mu_B$ of the isolated Mn
impurity. The most stable is the FM-ZB structure and not the RS
one showing a different behavior with respect to the
early-transition metal $\delta$-doping. As shown in Fig.
\ref{fig3} the FM-ZB is in reality a ferromagnetic semiconductor
while the FM-RS is a special case of spin-gapless semiconductor
since the top of the spin-up valence touches the bottom of the
spin-down conduction band exactly at the Fermi level.

\begin{figure}[t]
\includegraphics[width=\columnwidth]{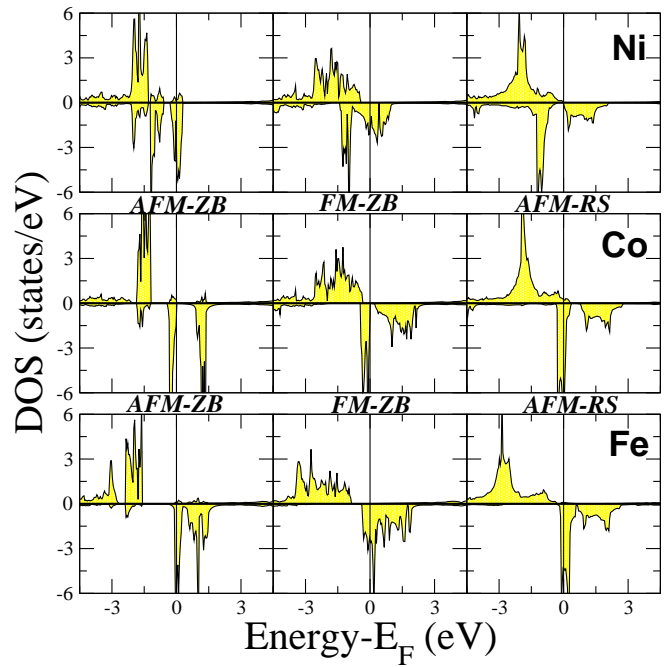}
\vskip -0.2 cm \caption{Similar to Fig. \ref{fig1} for the Fe, Co
and  Ni $\delta$-doped systems under study.} \label{fig4}
\end{figure}

As we substitute Fe(Co,Ni) for Mn we start populating also
spin-down states. and the atomic spin magnetic moments decrease by
about 1 $\mu_B$ as we move from Mn to Fe and then to Co and Ni
since we increase the number of valence electrons by one. In all
three Fe-, Co- and Ni- cases we were not able to converge the
FM-RS structure and the ground state was found among the two
magnetic ZB configurations as for the Mn-case (AFM-ZB for Fe and
Co and FM-ZB for Ni). As shown in Fig. \ref{fig4} the cases of Fe,
Co and Ni $\delta$-doping in the ground ZB structure should not be
stable since the Fermi level crosses intense spin-down picks.

To conclude using first-principles calculations we present
preliminary results on the magnetic ordering in the case of
transition-metal $\delta$-doping in cubic zinc-blende (ZB) and
rock-salt (RS) structures. The ZB is expected to represent also
the phenomena occurring in the ground wurtzite structure. When the
doping is of Fe, Co and Ni character we expect the resulting
structure to be unstable. Mn doping leads to stable ferromagnetic
coupling while in all three Ti, V and Cr doping-cases the
antiferromagnetic RS structure is favored. Mn-based systems are
magnetic semiconductors while the rest of the early transition
metal atoms (Ti, V, Cr) results in half-metallic behavior. Further
calculations using more dense grids in the reciprocal space as
well as calculations including the on-site Coulomb correlations
(Hubbard $U$) are programmed to be performed as the next step.
Especially on-site correlations should play a decisive role since
it has been already shown that even in bulk ZnO the electronic
properties are altered when the former are included,\cite{U} and
the effect is even larger in transition-metal doped
ZnO.\cite{Nayak,Gopal}

\textit{Acknowledgment} {Author acknowledges that this research
has been co-financed by the European Union (European Social Fund -
ESF) and Greek national funds through the Operational Program
"Education and Lifelong Learning" of the National Strategic
Reference Framework (NSRF) - Research Funding Program: THALES.
Investing in knowledge society through the European Social Fund.}

\end{document}